%% file: final.tex
\documentclass{article}

    \usepackage[final]{neurips_2020}

\input{shortcuts}

\title{Multi-Principal Assistance Games: \\ Definition and Collegial Mechanisms}

\author{%
   Arnaud Fickinger\\
  Department of EECS\\
  University of California, Berkeley\\
  \texttt{arnaud.fickinger@berkeley.edu} \\
   \And
   Simon Zhuang\\
   Department of EECS\\
   University of California, Berkeley \\
   \texttt{simonzhuang@berkeley.edu} \\
   \And
   Andrew Critch \\
  Department of EECS \\
    University of California, Berkeley \\
   \texttt{critch@berkeley.edu} \\
   \And
  Dylan Hadfield-Menell \\
  Department of EECS \\
    University of California, Berkeley \\
   \texttt{dhm@berkeley.edu} \\
   \And
  Stuart Russell \\
  Department of EECS \\
    University of California, Berkeley \\
   \texttt{russell@berkeley.edu} \\
}

\begin{document}

\maketitle

\begin{abstract}

We introduce the concept of a \emph{multi-principal assistance game} (MPAG), and circumvent an obstacle in social choice theory --- Gibbard's theorem --- by using a sufficiently ``collegial'' preference inference mechanism. % AC: it's collaboration-based in the sense that the humans have to get some of the utility themselves.
In an MPAG, a single agent assists $N$ human principals who may have widely different preferences.  MPAGs generalize \emph{assistance games}, also known as cooperative inverse reinforcement learning games.  We analyze in particular a generalization of apprenticeship learning in which the humans first perform some work to obtain utility and demonstrate their preferences, and then the robot acts to further maximize the sum of human payoffs.  
%thm:lessmanimulable --> 
We show in this setting that if the game is sufficiently \emph{collegial}
--- i.e., if the humans are responsible for obtaining a sufficient fraction of the rewards through their own actions --- then their preferences are straightforwardly revealed through their work.  
%thm:lessmanimulable -->
This revelation mechanism is non-dictatorial, does not limit the possible outcomes to two alternatives, and is dominant-strategy incentive-compatible.  
% AC: optional"
% Limitations and implications of the results are discussed.

\end{abstract}

% AC: Related work:
% other ways to overcome G–S; ours is not the only way, but an important way.

\section{Introduction}

The growing presence of AI systems that collaborate and coexist with humans in society highlights the emerging need to ensure that the actions of AI systems benefit society as a whole. This question is formalized as the \textit{value alignment} problem in the AI safety literature \cite{amodei2016concrete}, which emphasizes the need to align the increasingly powerful and autonomous systems with those of their human principal(s). However, humans are prone to misspecify their objectives which can lead to unexpected behaviors \cite{amodei2016concrete}; hence research in value alignment has focused on deriving preferences from human actions. In the body of research in value alignment and human robot interaction, the majority of the work involves scenarios with one human and one AI system. It is an appealing setting because the robot and the human share the same goal. Therefore, methods in this setting such as inverse reinforcement learning \cite{ng2000algorithms, abbeel2004apprenticeship, ramachandran2007bayesian}, inverse reward design \cite{hadfield2017inverse}, and LILA \cite{DBLP:journals/corr/abs-1906-10187} revolve around how an AI system can optimally learn the preferences of the human and apply these results to novel environments. Similarly, the human's incentive is to optimally teach the robot its own preferences. The combination of a learning AI system and a teaching human yields the \emph{assistance game} (also known as the cooperative inverse reinforcement learning game) \cite{hadfield2016cooperative}.

However, AI systems in the real world do not fit this \textit{one human, one AI} paradigm. Recommendation systems, autonomous vehicles, and parole algorithms do not exist in a vacuum---they often influence and are influenced by multiple humans. Hence, we consider a variation on assistance games where a robot acts on behalf of multiple humans, which we call the \emph{multi-principal assistance game} (MPAG). The key difference between this and the scenario with only one human is that, in general, different humans have different preferences, so it is impossible to align the AI to perfectly match the preferences of everyone. The problem of aggregating individual preferences for making collective decisions has been studied by economists and philosophers for more than two hundred years and constitutes the heart of social choice theory \cite{sen1986social}. 

Even with a given aggregation method, however, the inference process itself is challenged by the presence of selfish agents. While the robot acts to optimize the aggregate of utilities, each human acts to optimize their own utility. Therefore, unlike the single-principal assistance game, the multi-principal assistance game is no longer fully cooperative. This creates a problem for existing value alignment algorithms. These algorithms work under the assumption that the demonstrations and information provided are truly representative of the human's preferences. However, the misalignment between the AI system and each human's preferences yields a perverse incentive for the humans: can they misrepresent their preferences to gain a more desirable outcome? 

In this work we introduce the alignment problem of an AI system with multiple principals and establish a strong connection with results in computational social choice theory and mechanism design. We consider a subclass of MPAGs that generalizes apprenticeship learning \cite{abbeel2004apprenticeship}. In \textit{multi-principal apprenticeship learning}, the robot observes trajectories from multiple humans and then produces a trajectory that maximizes a social aggregate of the inferred rewards.

Our contributions are as follows:
\begin{itemize}
    \item We introduce the problem of learning from multiple strategic demonstrators with possibly wildly divergent reward and formulate an impossibility result in this context.
    \item We introduce an algorithm that manipulates feature-matching \textit{Learning From Demonstration} (LfD) algorithm in polynomial time, thus challenging the fact that computational hardness can be a barrier to manipulation in this context.
    \item We relate this problem to a real-world example, emphasizing the need to push research towards this direction.
    \item We propose a \textit{collegial} mechanism to circumvent the impossibility result in this context. Specifically, a collegial mechanism exploits the fact that when collected in the field, demonstrations can have consequences for their demonstrators that are independent of the behavior of the AI system.
\end{itemize}

% However, giving expert demonstrations by maximizing reward in isolation is often a suboptimal teaching algorithm, and there is research in how optimal teaching occurs in Markov decision problems \cite{Cakmak2012AlgorithmicAH}. In the multi-principal assistance game, especially, the socially optimal trajectory for a robot may not be individually optimal for any specific human, and hence may not be reflected in expert demonstrations. As our final contribution, we derive an efficient, incentive-compatible algorithm where humans are incentivized to share the full spectrum of their preferences. 

%\textcolor{red}{Our contribution is two-fold. Firstly, we bring a new perspective on the impossibility theorems in Social Choice Theory. More specifically, we show that inferring preferences by observing humans constitute a natural mechanism design. Secondly, we introduce and formalize the multi-agent alignment problem and show theoretically and experimentally how it is fundamentally harder than the single-agent counterpart}

\subsection{Related Work}
\paragraph{Value Alignment.}  %Inspired by the ability of young infants to infer goals by simply observing an adult's behavior, previous work have developed learning from demonstrations (LfD) methods to infer the goal without requiring the human to explicitly express it \cite{woodward1998infants} \cite{meltzoff1995understanding}. 
The need for AI systems to align with the preferences of humans  is well documented in AI safety literature \cite{amodei2016concrete}. A first line of work formulates goal inference as an inverse planning problem \cite{baker2007goal}. For example, Inverse Reinforcement Learning (IRL) computes a reward such that the observed trajectory is optimal in the underlying Markov Decision Process (MDP) \cite{ng2000algorithms} \cite{ziebart2008maximum}. A common assumption of inverse planning methods is that the robot does not influence the decision-making of the human. However, previous work has shown that the presence of a robot has a significant influence on humans \cite{robins2004effects} \cite{kanda2004interactive}. Furthermore, it has been shown that the robot can benefit from interacting with the human to infer the goal. For example, Hadfield-Menell et al. have shown that if we formulate goal inference as a game between the human and the robot, observing the optimal trajectory of the human is not a Nash equilibrium of the game in general \cite{hadfield2016cooperative}. Our work extends this idea to the multi-agent setting: we show that when a robot acts on behalf of multiple humans using LfD tools for single-agent alignment, the best strategy of the humans depends on the strategy of the other humans. This motivates the need for developing LfD tools specifically for multi-agent alignment.

\paragraph{Computational Social Choice.} Social Choice Theory (SCT) is a branch of Economics that studies the aggregation of individual preference towards a collective choice and encompasses many real-world scenarios like voting, fair allocation and auctions. A famous result of SCT is Gibbard's impossibility theorem which loosely states that any non-trivial\footnote{A process is non-trivial when it is neither dictatorial nor limiting the possible outcome to two options only.} process of collective decision is subject to manipulation \cite{gibbard1978straightforwardness}. Much effort in the Computational Social Choice and Mechanism Design communities has been focused on identifying situation where Gibbard can be circumvented and develop computational tool against manipulation \cite{brandt2016handbook}. A first line of work exploits the fact that there are some restrictions on the domain of preferences such that Gibbard doesn't hold anymore. Two widely studied domain restrictions are the single-peaked preferences in the voting literature \cite{black1958theory}, requiring the utility functions to be uni-modal, and the quasi-linear utilities in the auction literature \cite{groves1973incentives}, requiring money transfer between the users and the system to be applicable to the real world. Yet real-world demonstrations usually come from multi-modal reward function and we don't consider money transfer between the AI system and the users here, thus this restrictions are not applicable in our context. A second line of work exploits the fact that it might be computationally hard to manipulate a system. Yet we show that it is not hard to manipulate the value alignment methods we are considering in this paper by proposing an algorithm that computes a best-response in polynomial time. In this work we propose a natural way to circumvent Gibbard when learning from demonstration: collecting demonstrations that are meaningful for the demonstrator.

\paragraph{Cost of Lying.}
As we will see, collegial mechanisms incur a natural cost of lying to the demonstrator. Cost of lying has been introduced in many different scenarios, eg. guilt aversion \cite{battigalli2013deception}, altruism \cite{kerschbamer2019altruists} and reciprocity \cite{fehr2000fairness}. A widely studied model in economics is the model of partial verification, where the system can detect a lie when it is too far from the truth \cite{caragiannis2012mechanism}, in that case inflicting an infinite cost to the liar. Our model can be seen as a soft version of partial verification \cite{10.1145/2940716.2940795}. A line of work in the voting literature is interested in costly voting, where voters can pay more or less to express the degree of their preferences \cite{lalley2018quadratic}. Yet it is hard to link the utility to the willingness to pay, especially when voters have unequal wealth.

\paragraph{Learning from multiple demonstrators.}
Few works have been interested in learning a single-agent task from multiple demonstrators. Castro et al. introduce a maximum margin algorithm that exploit the fact that the multiple demonstrators have different known levels of expertise \cite{castro2019inverse}. Noothigattu et al. show that feature matching algorithms recover a good policy when the demonstrators are optimal with respect the a random perturbation of the same underlying reward \cite{noothigattu2020inverse}. In contrast, we show that feature matching algorithms are easily manipulable by a strategic demonstrator. None of these works consider strategic demonstrators.

\paragraph{Human-Robot Team} Robot evolving in a multi-human environment has already been studied by the Human-Robot Interaction community. Much work has focused on trust building and resource allocation \cite{claure2019reinforcement}. A common assumption is that the robot and the humans have a common payoff known to the robot. Our work generalize this setting to general-sum payoffs possibly unknown to the robot.

\section{Impossibility Result for Multi-Agent LfD Methods}

\subsection{Application of Gibbard's Theorem to LfD}
We consider a finite Markov Decision Process without reward $(S,A,P,\mu_0,T)$ where:
\begin{itemize}
    \item $S$ is a finite set of states.
    \item $A$ is a finite set of actions.
    \item $P: S \times A \times S \rightarrow [0,1]$ is a stochastic transition function.
    \item $\mu_0$ is a initial state distribution
    \item $T$ is a finite horizon
\end{itemize}
In IRL, an AI system observes an expert trajectory $\tau \in (S\times A)^T$ and computes a reward $R:S\times A \rightarrow \mathbb{R}$ that makes this trajectory optimal \cite{ng2000algorithms}. Apprenticeship Learning (AL) methods uses this reward as a proxy to compute a stochastic policy $\pi: S\times A \rightarrow [0,1]$ that best imitates the expert \cite{abbeel2004apprenticeship}. 

We propose \textit{multi-principal apprenticeship learning} as a generalization of AL. We suppose that there are $N$ demonstrators, each with a private reward function $R_i:S\times A \rightarrow \mathbb{R}$  and providing one trajectory $\tau_i$ to the AI system. The AI system observe all trajectories, compute a stochastic policy using an AL method for example and follows the policy to produce a trajectory $\tau_R \in (S\times A)^T$. The goal of the AI system is to maximize a social welfare function $W$ of the true rewards:
\begin{equation}
\begin{aligned}
\tau_R^* \in \argmax_{\tau} W(R_1(\tau),...,R_N(\tau))
\end{aligned}
\end{equation}
Social welfare functions are a heavily studied field, examples include the \textit{utilitarian} criterion $W_U(R_1,...,R_N) = \sum_h R_h$ \cite{6918520} and the \textit{egalitarian} criterion $W_E(R_1,...,R_N) = \min_h R_h$ \cite{zhang2014fairness,nace2008max}. In the remainder of the paper we consider the utilitarian criterion.

The process leading to $\tau_R$ can be represented by a stochastic function $g: ((S\times A)^T)^N \rightarrow \triangle((S\times A)^T)$. The objective of human $i$ is to lead the AI system towards a trajectory that maximize their own utility:
\begin{equation}
\begin{aligned}
\tau_i^* \in \argmax_{\tau_i} \mathbb{E} R_i(g(\tau_i, \tau_{-i}))
\end{aligned}
\end{equation}

Since $(S\times A)^T$ is a finite non-empty set, $g$ is a \textit{game form} as defined by Gibbard and we can apply his impossibility result for non-deterministic process \cite{gibbard1978straightforwardness}:
\begin{theorem} [Gibbard 1978]
	On the domain of versatile\footnote{A trajectory is versatile if the set of utility profile for which it is dominant has interior points.} trajectories, any straightforward\footnote{A straightforward mechanism induces a game where every player has a weakly dominant strategy. It is equivalent to say that it is not manipulable.} mechanism must be a probability mixture of mechanisms of two kind:
	\begin{itemize}
		\item Duple mechanisms, where the set of possible trajectories are restricted to two.
		\item Unilateral games, where one human gets to choose among a certain set of possible lotteries over trajectories.
	\end{itemize}
	\end{theorem}

Thus the only straightforward LfD mechanisms are not acceptable mechanisms. A first solution widely explored in the mechanism design literature would be to constraint the domain of preferences. Yet reward functions for real-world tasks can have various structure and be highly multi-modal, and we don't consider here money transfer between the demonstrators and the AI system. If we can't constraint the domain of preferences, we can hope that manipulating an AI system observing trajectories is computationally hard. Yet we challenge this hope in the following section by introducing an algorithm that manipulates feature matching algorithms, a widely used family of LfD algorithms, in polynomial time.

\subsection{A polynomial-time algorithm to manipulate feature-matching algorithms}

Current methods in learning from demonstrations are not adapted for dealing with multiple humans because they are easily manipulable. For example, if an AI system uses a feature matching algorithm on an aggregation of demonstrations coming from different humans, we can find a demonstration such that the robot policy is biased towards a single demonstrator in polynomial time. More specifically, we have the following result:

\begin{proposition}
Suppose that the AI system models the humans as noisily-optimal planners with linear features and computes the rewards that maximize the likelihood of the aggregated demonstrations \cite{ziebart2008maximum}. The best-response trajectory of a human can be computed in polynomial time.
\end{proposition}

To show that, we transform the manipulation problem into a least square problem. The objective of the AI system is the following:
\begin{equation}
\begin{aligned}
\omega^* & = \max_{\omega} P(\tilde{\tau}|\omega,\rho_0) \\
P(\tilde{\tau}|\omega,\rho_0)& = \prod_{i=1}^{N}\frac{e^{\phi(\tau^i)^T\omega}}{Z(\omega, \rho_0)} \\
Z(\omega, \rho_0) & = \sum_{\tau, s_0 \sim \rho_0} e^{\phi(\tau)^T\omega}
\end{aligned}
\end{equation}

where $N$ is the number of demonstrators, $\tilde{\tau} = (\tau_1,...,\tau_n)$ is the aggregate of trajectories and $\phi(\tau^i)^T\omega = \sum_{t=1}^{T}\phi(s_t^i)^T\omega$ is the cumulative return of $\tau^i$ under reward $\omega$.

By taking the gradient to zero, we see that this concave objective is maximized when the expected feature count of the computed reward’s optimal policy is equal to the empirical feature count of the aggregated demonstrations:
\begin{equation}
\begin{aligned}
\mathbb{E}_{\tau \sim \pi^*(w^*), s_0 \sim \rho_0}(\phi(\tau)) = \frac{\sum_{i=1}^{N} \phi(\tau^i)}{N}
\end{aligned}
\end{equation}
This gives the strategic demonstrator a simple procedure to bias the system towards their own interest: they give a demonstration such that the empirical feature count of the aggregated demonstrations is the closest possible to their own policy’s expected feature count. Formally, this translates into the following least squares objective:
\begin{equation}\label{eq:3}
\begin{aligned}
\tau^*_i(\tau_{-i}) = \min_{\tau_i} \norm{\phi(\tau_i) - (N\mathbb{E}_{\tau \sim \pi^*(w_i), s_0 \sim \rho_0}(\phi(\tau) )- \sum_{j\neq i}\phi(\tau_j) )}^2_2
\end{aligned}
\end{equation}
Concretely, a strategic demonstrator will act like they like (dislike) something more than they actually do to push the AI system towards (away from) a particular outcome.

\begin{figure}[h]
\centering
  \begin{subfigure}[b]{0.45\textwidth}
    \includegraphics[width=\textwidth]{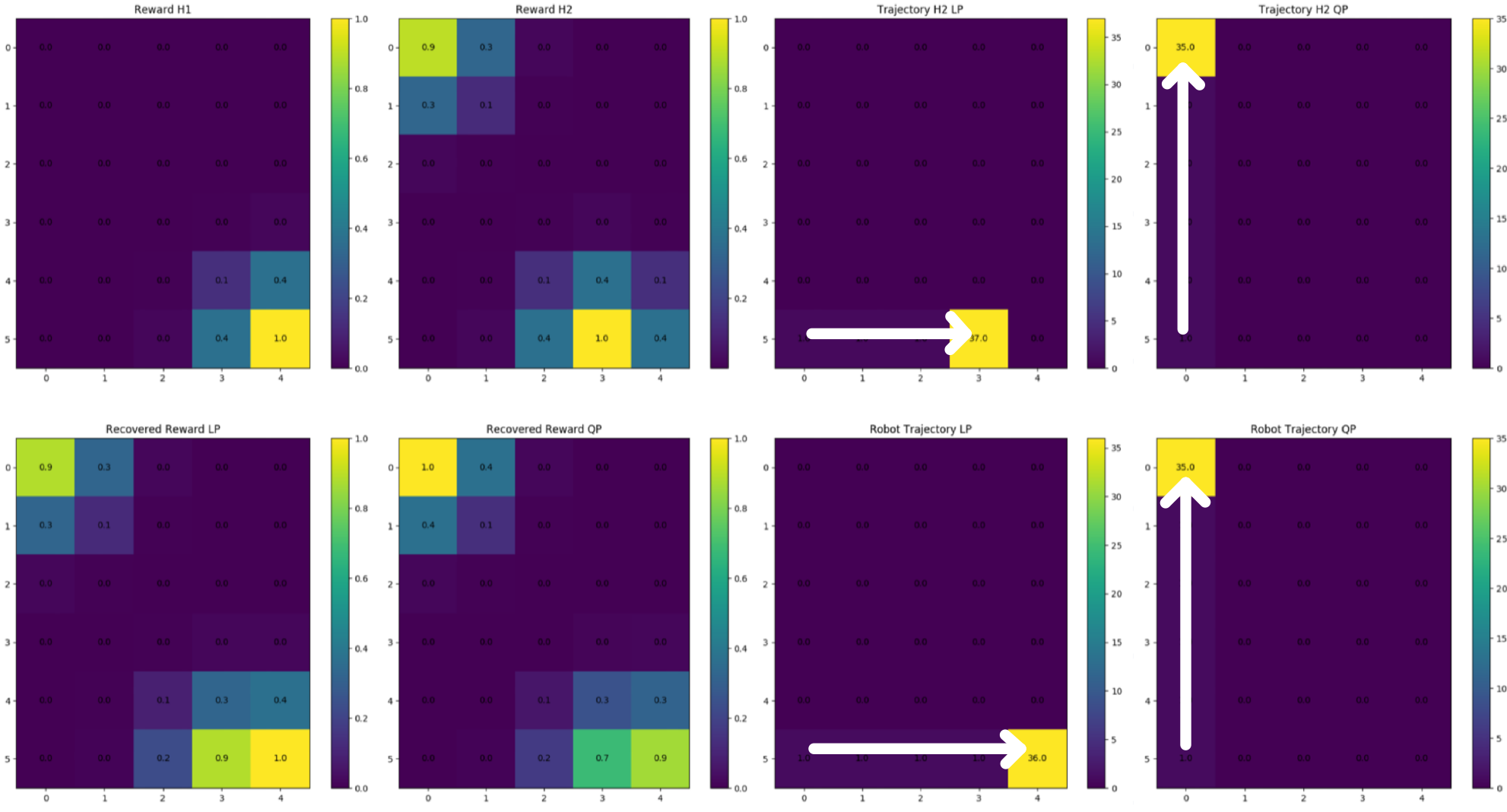}
    \caption{First row: $R_1$; $R_2$; $H_2$'s honest trajectory; $H_2$'s strategic trajectory. Second row: Recovered reward when $H_2$ is honest; Recovered reward when $H_2$ is strategic; Robot's trajectory when $H_2$ is honest; Robot's trajectory when $H_2$ is strategic ($H_1$ is always honest).}
    \label{fig:1}
  \end{subfigure}
  \begin{subfigure}[b]{0.45\textwidth}
    \includegraphics[width=\textwidth]{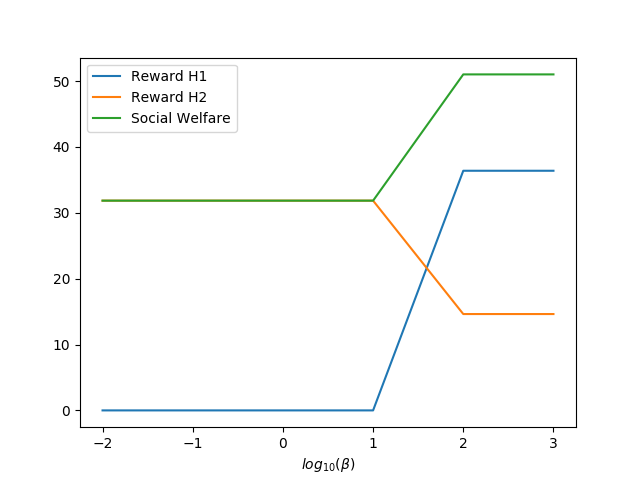}
    \caption{Return of $H_1$ and $H_2$ and social welfare from the robot's trajectory against $\beta$ ($H_1$ is always honest).}
    \label{fig:2}
  \end{subfigure}
\end{figure}

An efficient way to solve this problem is to find the occupancy measure that minimizes the following constrained least squares problem:
\begin{equation}\label{eq:opt}
\begin{aligned}
&\min_{\rho^t_{s, a}} & ||\sum_{s, a, t} \rho^t_{s, a} \phi(s)-  (N\mathbb{E}[\phi | w]-\sum_{j \neq i}\phi(\tau_j))||^2 \\
&\text{subject to} & \sum_{a} \rho^{t+1}_{s', a} =  \sum_{s, a} P(s, a, s') \rho^t_{s, a} \, \, \forall s', \forall t\in [0, T-1]\\
&&\sum_{a} \rho^0_{s, a} = \mu_0[s] \, \, \forall s
\end{aligned}
\end{equation}
This is a linearly constrained least squares problem solvable in polynomial time. We can compute the best-response trajectory directly from the occupancy measure in linear time.
We test our solver in a gridworld environment (see Fig. \ref{fig:1}) and observe that:
\begin{itemize}
    \item The best-response trajectory can be computed in very short time, thus computation time is not a barrier to manipulation.
    \item The best-response trajectory is not what the agent would have picked in isolation and depend on other humans' trajectories. In other words, this LfD method is not straightforward in general, in accordance with Gibbard's theorem. 
\end{itemize}

% \begin{figure}[h]
% \centering
% \begin{minipage}{.45\textwidth}
% %   \centering
%   \includegraphics[width=1\linewidth]{image(1).png}
%   \caption{First row: $R_1$; $R_2$; $H_2$'s honest trajectory; $H_2$'s strategic trajectory. Second row: Recovered reward when $H_2$ is honest; Recovered reward when $H_2$ is strategic; Robot's trajectory when $H_2$ is honest; Robot's trajectory when $H_2$ is strategic ($H_1$ is always honest).}
% %   \label{fig:test1}
% \end{minipage}%
% \begin{minipage}{.45\textwidth}
% %   \centering
%   \includegraphics[width=0.8\linewidth]{collegialq.png}
%   \caption{Return of $H_1$ and $H_2$ and social welfare from the robot's trajectory against $\beta$ ($H_1$ is always honest).}
% %   \label{fig:test2}
% \end{minipage}
% \end{figure}

% \begin{figure}[h]
% \centering
% \includegraphics[width=0.5\textwidth]{}
% \includegraphics[width=0.4\textwidth]{collegialq.png}
% \caption{Manipulating a Multi-Agent Alignment IRL Method using a QP in a 2D $5 \times 6$ Gridworld Environment with a 3D feature space.  First row: True reward of humans 1 and 2; State visitation count of optimal (resp. best-response) trajectories of human 2 (the initial state is in the bottom left-hand corner). Second row: Recovered rewards using IRL on the aggregate of first human's optimal and second human's optimal (resp. best-response) trajectories; Optimal robot trajectories in the MDP induced by these rewards.}
% \label{fig:1}
% \end{figure}

\subsection{An Example of Real-World Misalignment: Microsoft Tay}

% This objective illustrates what the Tay's pranksters were trying to do: they were using extremely racist words to push the total empirical feature count towards more racist features than used in normal conversations.

In 2016, Microsoft released a Twitter chatbot, Tay, designed to learn to converse via tweets. It took less than 24 hours for a group of prankster users to train Tay to mix racist comments into its discourse. Tay had at least three conceptual problems:
\begin{enumerate}
    \item Manipulable inputs. Tay was not trained on chat logs ‘in the wild’; it was trained by humans who knew there was a system that could be manipulated to achieve goals outside its intended purpose.
    \item User/creator misalignment. TAY’s creators primarily wanted TAY to imitate a normal person, not entertain people (although entertainment was useful to gain more engagement and data).  By contrast, its users primarily wanted to be entertained.  This means there was a misalignment between the creators as principals and the users as principals.
    \item User/user misalignment. Prankster users wanted offensive entertainment, presenting a misalignment between different groups of the users.  The pranksters were able to increase TAY’s level of racism to an unusual degree using:
    \begin{enumerate}
        \item extreme features (highly racist inputs)
        \item extreme numbers of inputs 
\end{enumerate}
\end{enumerate}

In this work, we are concerned primarily with problems analogous to problem 1 and problem 3(a). Problem 1 raises the question of how to elicit honest responses, i.e., incentive compatibility. Problem 3(a) raises the question of users exaggerating their preferences to move the value of a ‘compromise’ policy closer to their desired policy, which is a kind of manipulability for many learning procedures that we have recovered in the previous subsection by computing the best-response to a feature-matching LfD algorithm (see Equ. \ref{eq:3}). As we will see in the next section, these problems can both be mitigated by requiring demonstrations to have natural consequences outside of the AI’s policy, thereby eliciting more ‘normal’ behavior from demonstrators.

\section{Circumventing Gibbard's Theorem with Collegial Mechanisms}
\subsection{Exploiting the Consequential Nature of Real-World Demonstrations}

Even if it is not hard to compute what trajectory would bias a system towards an individual interest, there are scenarios in real life where a human would prefer to stay honest. This is due to the fact that when collected in the field, demonstrations might have consequences for their demonstrators that are independent of the behavior of the AI system. 
In our setting, this can be formulated using the same objective with an additional term, the direct reward the human will get by choosing an action with a coefficient $\beta$ that quantifies the importance that the human puts on the direct consequences of its demonstration relatively to the robot’s action:
\begin{equation}
\begin{aligned}
\tau_i^* \in \argmax_{\tau_i} \beta R_i(\tau_i) + \mathbb{E} R_i(g(\tau_i, \tau_{-i}))
\end{aligned}
\end{equation}

To obtain a clear bound on $\beta$ we suppose that the reward function are integer-valued: $R_i: S \times A \rightarrow \mathbb{N}$ and that there is $M \in \mathbb{N}$ such that: $\forall i, \forall (s,a) \in S \times A, R_i(s,a) \leq M$. Notice that the Gibbard's impossibility result stated in the previous section still holds when we take integer-valued reward functions with a fixed upper-bound. 
We have the following result:
\begin{proposition}\label{thm:stra1}
If $\beta>M$, then every mechanism is straightforward.
\end{proposition}
This result shows that we can circumvent Gibbard's theorem by looking for situation where demonstrations are the most meaningful, incurring a natural cost of lying for the demonstrator.

A similar bound can be obtained for real-valued reward functions. For every human $i$, we define $R^*_i = \max_{s,a} R_i(s,a)$, we assume that there is $(s,a)$ such that $R(s,a)<R^*_i$ and we define $\gamma_i = \min_{(s,a)}\{R^*_i-R_i(s,a): R(s,a)<R^*_i\}$. We also define $\gamma = \min_i \gamma_i$. Since we consider a finite MDP with a finite number of humans we have $\gamma>0$. We have the following result:
\begin{proposition}\label{thm:stra2}
If $\beta>\frac{M}{\gamma}$, then every mechanism is straightforward.
\end{proposition}

Even when $\frac{M}{\gamma} \rightarrow \infty$ and $\beta<\frac{M}{\gamma}$, collecting meaningful demonstrations can significantly reduce the manipulability of a mechanism. To see that, we consider a plurality voting system with random tiebreak with 3 voters and 3 alternatives and compare the proportion of manipulable profile when $\beta = 0$ and $\beta = 1$ for the utilities domain $\{R:\{1,2,3\} \rightarrow [0,1], R(1)+R(2)+R(3)=1\}$ under which Gibbard's theorem still holds and such that $\gamma \rightarrow 0$. Using a geometric argument on the 2-simplex we show that:
\begin{proposition} \label{thm:stra3}
In a system using plurality voting with random tiebreak with 3 voters and 3 alternatives, $\frac{1}{3}$ of the simplex is manipulable\footnote{We say that a utility function is manipulable when the associated best response strategy depends on the strategy of the other humans.} when $\beta=0$ while only $\frac{1}{9}$ of the simplex is manipulable when $\beta=1$.
\end{proposition}
We can efficiently compute the best-response trajectory when $\beta>0$ by adding a term to the objective of the previous optimization problem (see Equ. \ref{eq:opt}):
\begin{equation}
\begin{aligned}
&\max_{\rho^t_{s, a}} & \beta \sum_{s, a, t} \gamma^t \rho^t_{s, a} \phi(s)^T w - ||\sum_{s, a, t} \rho^t_{s, a} \phi(s)-  (N\mathbb{E}[\phi | w]-\sum_{j \neq i}\phi(\tau_j))||^2 \\
&\text{subject to} & \sum_{a} \rho^{t+1}_{s', a} =  \sum_{s, a} P(s, a, s') \rho^t_{s, a} \, \, \forall s', \forall t\in [0, T-1]\\
&&\sum_{a} \rho^0_{s, a} = \mu_0[s] \, \, \forall s
\end{aligned}
\end{equation}
We recover a regularized dual of the linear program formulation for finite-horizon discounted Markov Decision Process \cite{puterman2014markov}. 

We plot the social welfare obtained by the robot's policy in our gridworld setting against the importance that the strategic demonstrator put on their demonstration (see Fig \ref{fig:2}). We observe that when $\beta$ is higher than 100, $H_2$ is incentivized to be honest and the social welfare increases significantly.

Thinking back about our real-world example, if the Tay bot had been reading from people’s work account instead of anonymous Twitter feed, the problem would not have occurred, since there’s a greater negative utility to the human for providing profane examples in the former case.

\subsection{Towards Efficient Mechanisms with Collaboration beyond Demonstrations}

So far we have considered a subclass of MPAGs where the AI system is passively observing the humans. Although it enables a clear comparison with the social choice theory, it is arguably not the best way to learn human values. A challenge of learning from multiple demonstrators is that demonstrations give only one mode of the reward function. Yet to maximize the social welfare we certainly need more information: there is cardinal utility profile such that the social maximizing action is sub-optimal with respect to each of the individual utilities\footnote{Consider the utility profile $\{(0.6,0.4,0), (0,0.4,0.6)\}$ in a stateless MDP with 3 actions and 2 humans.}. 

In this section we widen the considered class of MPAGs to yield an approximately efficient\footnote{A mechanism is efficient if it maximizes the social welfare.} mechanism. Specifically, we assume that the AI system learns human values through a human-robot collaborative task. Consequently, the AI system has an influence on the utility the human get when demonstrating their preferences. 

We obtain a non-trivial asymptotic worst case bound on the social welfare in a simplified model of human-robot collaboration. We consider a stateless sequential setting where at each time step, the robot can choose one human (among N humans) to collaborate with. During the collaboration, the human chooses one action (among M actions) and at the end of the step, the robot chooses weather the human get the associated reward. The robot chooses at which time step to stop and then chooses an action. We wish to maximize the social welfare of this action.

A robot mechanism $\mathcal{M}$ is given by a human selection criterion, a reward allocation criterion, a stopping time criterion and an action selection criterion. We define the distortion\footnote{The distortion is a notion introduced in the ordinal voting literature.} of the robot's mechanism as:
\begin{equation}
\begin{aligned}
\triangle(\mathcal{M}) = \max_{R}\frac{\max_a \sum_h R_h(a)}{\mathbb{E}\sum_h R_h(a_{\mathcal{M}}(R))}
\end{aligned}
\end{equation}
We propose a mechanism that achieves a non-trivial asymptotic distortion (see Algorithm \ref{algo:1}). In broad outline, the robot chooses a human and allocates reward only if the human did not choose the action before. Periodically, the robot chooses a random action with probability $1-\frac{1}{2^{\frac{1}{M}}}$.

Exploiting the fact that humans plan with a discount factor strictly less than 1 and using recent tools from the ordinal voting literature\cite{BOUTILIER2015190}, we obtain the following bound:
\begin{proposition} \label{thm:distortion}
 $\triangle(\mathcal{M}) = O(\sqrt{M\log M})$
\end{proposition}

\section{Conclusion}
In this paper, we explore an area of concern in the study of AI alignment---ensuring that AI systems are designed so that humans agents are incentivized to interact with AI systems in a ``honest" way.  We view our main contributions as follows:
\begin{itemize}
    \item Propositions \ref{thm:stra1}, \ref{thm:stra2} and \ref{thm:stra3} show that collegial preference inference can yield numerous desirable properties including incentive-compatibility.
    \item Proposition \ref{thm:distortion} reveals the asymptotic performance of a mechanism coupling collegial preference inference with human-robot collaboration.
\end{itemize}

These results appear to be reasons for optimism in the domain of mechanism design for multi-principal assistance games.  Meanwhile, the overall problem of preventing manipulative behavior in multi-human AI systems is open and presents many opportunities for further work. Our methods are applied to fairly simple problems; there exists a need to generalize these results to more general theoretical settings and more complicated situations in the real world.

% Problems and solution are demonstrations?

% LfDemo introduce both prob and solutions that are outside of existing work in  comp social choice

% Thus designing a mechanism that circumvents GS without money in our setting amount to find a way to collect demonstrations that are meaningful for the human. Furthermore our objective enable to shift the too restrictive view of GS, add a new dimension to GS. 
% Thinking of the fact that demonstrations can be directly consequential is the key to design good mechanism among humans.

% Thus, a mechanism that circumvents G–S can be achieved whenever there is a way to collect demonstrations have sufficiently strong natural consequences for the human outside of the AI system’s behavior.  A key objective of this paper is to convey this principle.

\bibliography{final}
\bibliographystyle{unsrt}
\newpage
\begin{algorithm}[h]
\caption{Approximately Efficient Mechanism}
 \label{algo:1}
\SetAlgoLined
\SetKwInOut{Input}{Input}
\SetKwInOut{Output}{Output}
\Input{Number of humans $N$; Number of arms $M$}
\Output{Robot action $a_R$}
\For{$h\in [1,N]$}{
$Actions[h] \longleftarrow [1,M]$
}
\For{$a\in [1,M]$}{
$Score[a] \longleftarrow 0$
}
\For{$t\in [1,M]$}{
\For{$h\in [1,N]$}{
Let human $h$ choose an action $a$

\If{$a \in Actions[h]$}{
Execute $a$

$Scores[a]+=\frac{1}{t}$

Remove $a$ from $Actions[h]$
}
}
With probability $1-\frac{1}{2^{\frac{1}{M}}}$, return an arm sampled uniformly on $[1,M]$.
}
Return arm $a$ with probability $\frac{Score[a]}{\sum_{a'}Score[a']}$

\end{algorithm}

\end{document}

%% file: shortcuts.tex
\usepackage{amsmath}
\usepackage{amssymb}
\usepackage{amsthm}

\newcommand{\norm}[1]{\left\lVert#1\right\rVert}

\newtheorem{theorem}{Theorem}

\newtheorem{proposition}{Proposition}

\usepackage[utf8]{inputenc} % allow utf-8 input
\usepackage[T1]{fontenc}    % use 8-bit T1 fonts
\usepackage{hyperref}       % hyperlinks
\usepackage{url}            % simple URL typesetting
\usepackage{graphicx}
\usepackage{listings}
\usepackage{subcaption}
\DeclareMathOperator*{\argmax}{arg\,max}

\usepackage{caption}
\usepackage[ruled,vlined]{algorithm2e}